\documentclass[%
 reprint,
 superscriptaddress,
 amsmath,amssymb,
 aps,
]{revtex4-2}

\usepackage{graphicx}
\usepackage{dcolumn}
\usepackage{bm}

\usepackage{subfigure}
\usepackage{array}
\usepackage{booktabs}
\usepackage{multirow}
\usepackage{siunitx}
\usepackage[utf8]{inputenc}
\usepackage[T1]{fontenc}
\usepackage{mathptmx}
\usepackage{etoolbox}
\usepackage{hyperref}
\begin{document}

\preprint{APS/123-QED}
\UseRawInputEncoding

\title{Experimental demonstration of dark current mitigation by an over-inserted plug in a normal conducting VHF gun}

\author{X.H. Wang}
\affiliation{University of Shanghai for Science and Technology, Shanghai, China}
\affiliation{Shanghai Institute of Applied Physics, Chinese Academy of Sciences, Shanghai, China}
\affiliation{Shanghai Advanced Research Institute, Chinese Academy of Sciences, Shanghai, China}
    
\author{G. Shu}
 \email{shuguan@zjlab.ac.cn}
\affiliation{Zhangjiang Laboratory, Shanghai, China}

\author{H. Qian}
 
\affiliation{Zhangjiang Laboratory, Shanghai, China}
	
\author{X. Li}
\affiliation{Shanghai Advanced Research Institute, Chinese Academy of Sciences, Shanghai, China}
	
\author{Z. Liu}
\affiliation{Zhangjiang Laboratory, Shanghai, China}
	
\author{Z. Jiang}
\affiliation{Shanghai Advanced Research Institute, Chinese Academy of Sciences, Shanghai, China}
	
\author{H. Meng}
\affiliation{Shanghai Advanced Research Institute, Chinese Academy of Sciences, Shanghai, China}

 \author{C. Xing}
\affiliation{Zhangjiang Laboratory, Shanghai, China}

 \author{Q. Zhou}
\affiliation{Zhangjiang Laboratory, Shanghai, China}

  \author{H. Deng}
  \email{denghx@sari.ac.cn}
\affiliation{Shanghai Advanced Research Institute, Chinese Academy of Sciences, Shanghai, China}

\date{\today}

\begin{abstract}
The room temperature continuous wave (CW) very-high-frequency (VHF) gun is one of the candidates for the electron gun of the high-repetition-rate free-electron lasers (FELs). The VHF gun operates with a cathode gradient of $\rm\sim$ 20 MV/m and an accelerating voltage of $\rm\sim$ 750 kV. The gun dark current emission leads to beam loss along the FEL machine, therefore is a critical parameter for the performance of the CW gun. In this paper, we presents a systematic study of the dark current reduction of the VHF gun, including cathode region optimizations, dark current tracking simulations and measurements. Over-inserted cathode plugs were tested in two VHF guns of different acceleration gap sizes, and both demonstrated significant dark current reduction ratios of more than two orders of magnitude.
\end{abstract}

\maketitle

\section{\label{sec:level1}Introduction}

Free electron lasers (FELs)  \cite{5,6,2,1,4,3,7,8,2013Using,9,2012Proposal} have shown great potentials for driving the frontiers of science and technology. To meet the demands of megahertz (MHz) repetition rate FELs, the electron gun should operate in the CW mode. Superconducting RF (SRF) guns \cite{10}, normal conducting (NC) VHF guns \cite{11,34,35} and direct-current (DC) guns \cite{12} are available electron guns for CW XFEL. The VHF gun adopts the mature and reliable normal conducting RF technologies which can provide a cathode gradient of 20 $\rm-$ 30 MV/m and a voltage of $\rm\sim$ 750 kV. Compared to both DC gun and SRF gun, VHF gun has a higher photoemission electric field at cathode, which is critical for maximizing initial beam brightness. The compatibility with the high quantum efficiency (QE) photocathode (e.g. $\rm Cs_{2}Te$), the stability of high RF power operation and the beam performance were demonstrated via APEX experiments. It was adopted as the electron gun for Linac Coherent Light Source \uppercase\expandafter{\romannumeral2} (LCLS-\uppercase\expandafter{\romannumeral2}) \cite{13}, Shanghai High repetition rate XFEL and Extreme light facility (SHINE) \cite{14}and Shenzhen Superconducting Soft-X-Ray Free Electron Laser ($\rm S^{3}FEL$)\cite{shenzhen}.

Higher cathode gradients and accelerating voltages in the gun are always pursued to achieve a higher beam brightness. The field emission, which increases exponentially with the electric field, is stronger as well. Partial field emission electrons can be accelerated and transmitted to the downstream beamline with a proper RF phase, forming dark current. The dark current induced beam loss might be a significant issue in the high repetition rate FEL facility, such as beam loss heating of the cryogenic system, beam loss radiation damage of electronics, undulator magnet degradation etc.
    
The field emission depends on several factors \cite{SF}, including the surface electric field, material work function, surface imperfections, such as contaminations and micro-tips. Several effective methods have been found to reduce the dark current in RF guns. The L band $\rm 5^{th}$ generation RF gun (named gun 5.1) at the Photo Injector Test facility at DESY in Zeuthen (PITZ) utilized elliptical plug hole corner instead of the round corner in the previous $\rm 4^{th}$ generation RF gun (named gun 4.2). The peak electric field on the hole was reduced by $\sim$ 14\% and the dark current was reduced by a factor of 3 to 5 compared to gun 4.2 \cite{18}. Experiments at PITZ shows electron guns treated with the $\rm CO_{2}$ dry-ice cleaning \cite{19} exhibited a dark current reduction by an order of magnitude compared to those rinsed with high-pressure water \cite{20,21}. Some gun experiments also showed that dark current may decrease during the RF conditioning process \cite{22}. The dark current can be controlled by using collimators \cite{23}, fast kickers \cite{25,26}, and chicanes \cite{27}. 
  
The experiments with the APEX gun revealed that the surface polishing and $\rm CO_{2}$ cleaning can achieve a reduction in dark current from 350 nA to 0.1 nA \cite{28}. The VHF gun at LCLS-\uppercase\expandafter{\romannumeral2} had a very low dark current of $\sim$10 nA at the beginning of its operation. After 700 hours of operation, the dark current increased to 2.5 $\rm \mu$A at the nominal cathode gradient (20 MV/m) and voltage (750 kV) \cite{29,30}. Then, the gun voltage was reduced to 650 kV and the dark current was $\sim$ 3.5 $\rm \mu$A, a collimator was installed in the low energy beam line to reduce dark current \cite{31,32}. A similar phenomenon was observed in the SHINE gun, and a collimator was also used \cite{14,33}.

The dark current of the VHF gun is a potential limitation for the injector performance in the SRF linac based FELs. The gun gradient is forced to decrease to control the dark current within an acceptable range, imposing a negative impact on beam quality. In this paper, a systematic dark current study of the VHF gun is presented. Experimental demonstration of the dark current reduction by an over-inserted cathode plug is achieved. This paper is organized as follows. In Section \ref{sec:2}, we provide a brief introduction to two VHF guns under dark current studies. Section \ref{sec:3} highlights the strategies employed to mitigate dark current through physical design, machining and surface processing. Section \ref{sec:4} presents the dark current measurements of a normal-inserted cathode plug. In Section \ref{sec:5}, we track the trajectories of the electrons emitted from the cathode vicinity and calculate the transmission ratio from the emission point to the Faraday cup (FC). The detailed simulations with various plug insertion depths are presented. Section  \ref{sec:6} presents the dark current experimental results of the over-inserted plugs in two VHF guns. The influence of the over-inserted plug on the beam dynamics is discussed. Section VII concludes and discusses this research work.

\section{\label{sec:2}VHF GUNS under dark current studies }

The dark current studies were conducted on two VHF guns: the SHINE VHF gun and ZJLAB/SARI VHF gun. 
\begin{table}[ht]
\centering 
\caption{\label{tab:table1}~Main RF parameters of SHINE and ZJLAB/SARI VHF guns.}
\begin{tabular}{lcc} 
\toprule
\textbf{Parameter} & SHINE gun & ZJLAB/SARI gun \\ \midrule 
Frequency [MHz] &\multicolumn{2}{c}{216.667} \\
Accelerating gap [mm] &30 &40 \\
Unloaded quality factor ($Q_0$) &33717 &32426 \\
Power coupling factor &\multicolumn{2}{c}{1.0} \\
Shunt impedance [M$\Omega$] &8.34 &7.89 \\
Cathode gradient [MV/m] &26 &19.8 \\
 Cavity voltage [kV] &750 &750 \\
 RF power [kW] &68 &72 \\
 Peak surface E [MV/m] &32.0 &22.6 \\
 Max. surface loss density [W/cm$^2$] &21.3 &24.5 \\ 
\bottomrule 
\end{tabular}
\end{table}
The VHF gun installed in the SHINE injector was developed by Tsinghua University. An APEX-type single-cell cavity was designed for CW operation at 216.667 MHz, which is the sixth subharmonic of the 1.3 GHz superconducting linac frequency. The accelerating gap is 3 cm in order to enhance the cathode gradient. During the high power conditioning, 75 kW CW power was fed into the gun, resulting in a cathode gradient of 27 MV/m and a gun voltage of 780 kV. More details about the gun development can be found in Ref. \cite{14}.

\begin{figure}
    \centering
    \includegraphics[width=6.8cm]{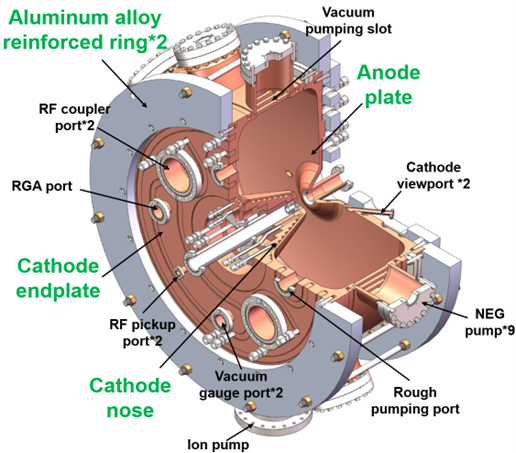}
    \caption{\label{fig1} ~Mechanical layout of ZJLAB/SARI VHF gun.}
    \label{fig1}
\end{figure}

An R $\rm\&$ D program for the VHF gun was initiated by Zhangjiang Laboratory (ZJLAB) and Shanghai Advanced Research Institute (SARI), Chinese Academy of Sciences to develop a mA-scale average current high brightness electron source \cite{liu1,36,liu2}. The ZJLAB/SARI VHF gun also serves as a backup solution for the SHINE injector. The ZJLAB/SARI gun, shown in Fig. \ref{fig1}, is composed of an anode plate, a cathode endplate, a cathode nose and two aluminum alloy reinforced rings. Two cathode plug viewports and four tuners are installed on the anode plate. The RF couplers, pickups, vacuum gauges, residual gas analyzer (RGA) and rough pumping valve are installed on the cathode endplate. The vacuum pumps are located around the side wall, including nine Non Evaporable Getter (NEG) pumps and one ion pump. The stainless steel (SS) vacuum pumping chamber in APEX gun, which has low thermal conductivity, is replaced by the cylindrical copper tubes on the sidewall. This modification reduces the gun's thermal equilibrium time and degassing surfaces. The cathode gradient is 19.8 MV/m and the gun voltage over a 4 cm accelerating gap is 750 kV. The main RF parameters of the SHINE and ZJLAB/SARI VHF guns are presented in Table \ref{tab:table1}. 

\section{\label{sec:3}EFFORTS IN THE DARK CURRENT REDUCTION}

Dark current control is a critical consideration that influences the physical design, machining, surface treatment and high RF power operation of the VHF gun. Previous studies indicate that the field emissions from the cathode vicinity are the major sources of the dark current. Since the field emission increases exponentially with the surface electric field, it is essential to optimize the geometries in cathode vicinity to reduce the surface electric field.

\begin{figure*}
    \centering
    \includegraphics[width=1\linewidth]{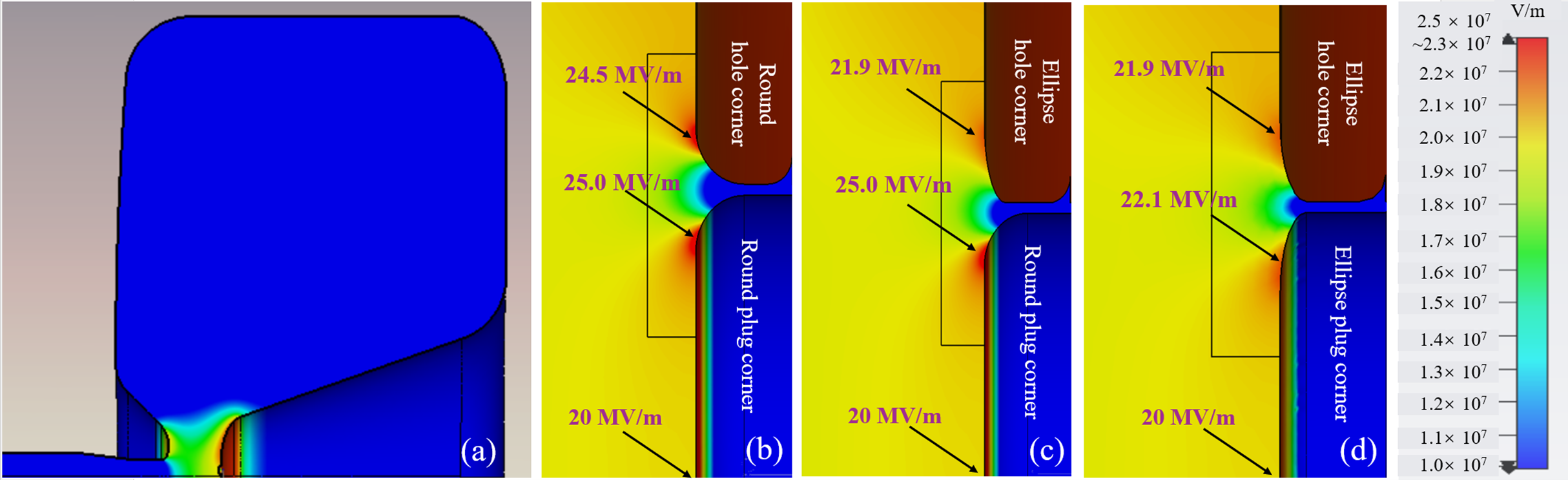}
    \caption{\label{fig2.png}~(a) Cross-sectional electric field distribution, the cathode gradient is normalized to 20 MV/m. Electrical field in the cathode vicinity with (b) Round corners, (c) Round plug corner and elliptical hole corner, (d) Elliptical corners.}
    \label{fig:enter-label}
\end{figure*}

There are three shape combinations for the cathode plug hole and the cathode plug, e.g. round hole corner and round plug corner in the LCLS-\uppercase\expandafter{\romannumeral2} VHF gun; elliptical hole corner and round plug corner in the PITZ gun 5.1; elliptical hole corner and elliptical plug corner in the SHINE VHF gun. We calculated the electric field distribution of these combinations by CST MICROWAVE STUDIO (CST MWS) 2022 \cite{CST}, as shown in Fig. \ref{fig2.png}. With an elliptical profile, the peak electric field on the cathode plug and plug hole are reduced by 11.7\% and 10.8\%, respectively, compared to the round corner. The ZJLAB/SARI VHF gun ultimately utilized the combination shown in Fig. \ref{fig2.png}(d).

\begin{figure}
    \centering
    \includegraphics[width=5cm]{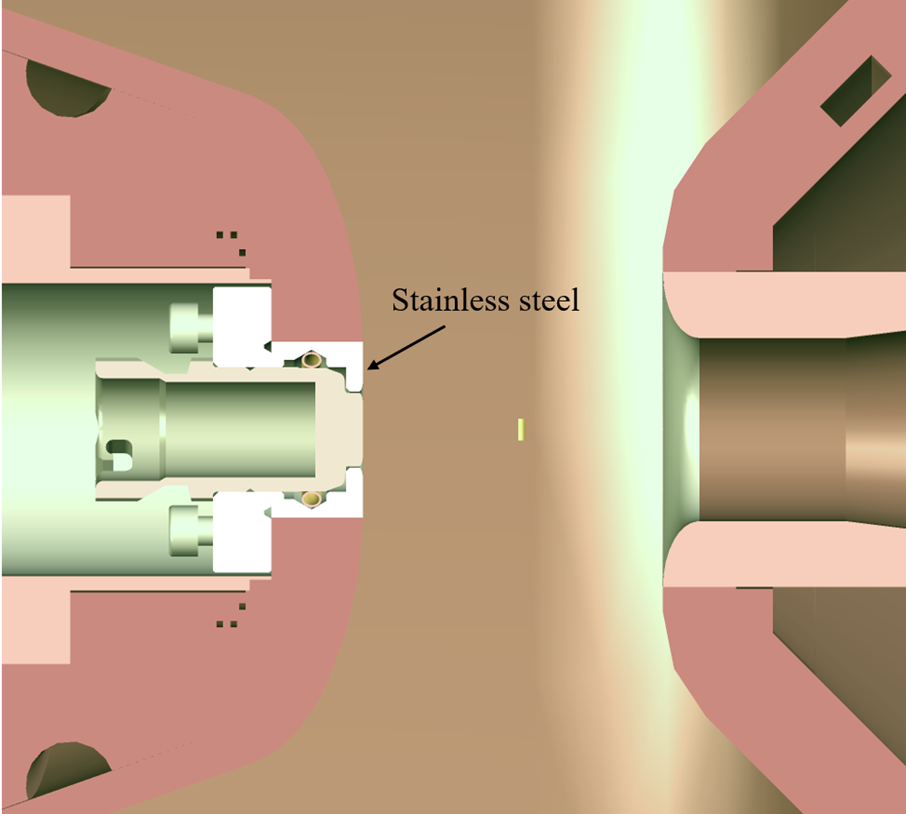}
    \caption{\label{fig3.png}~Cathode plug hole uses the SS 316L as the material.}
    \label{fig:enter-label}
\end{figure}

The SS cathode and anode nose inserts were proposed to reduce dark current in Ref. \cite{37}, due to its better chemical and mechanical robustness than copper. Consequently, we adopted the SS plug hole design in the ZJLAB/SARI VHF gun, as shown in Fig.  \ref{fig3.png}.

\begin{figure}
	\centering
	\subfigure[~]{\label{fig4.a.png}\includegraphics[width=6.8cm]{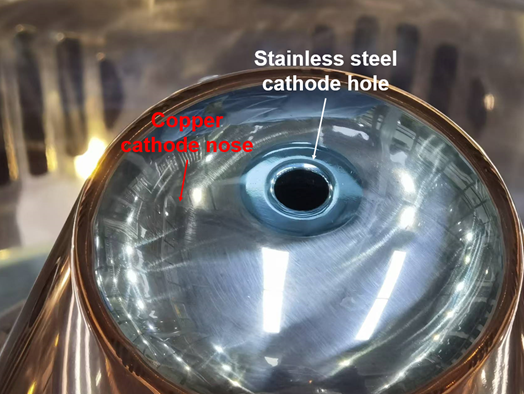}}\\
	\subfigure[~]{\label{fig4.b.jpg}\includegraphics[width=3cm]{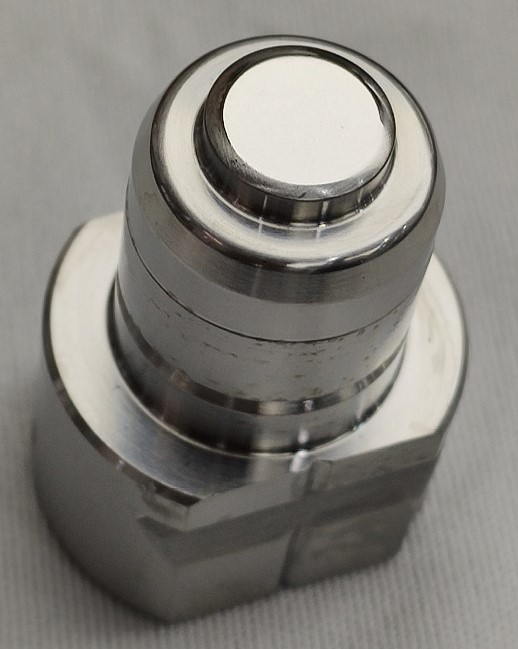}}
	\caption{~The cathode nose (a) and cathode plug (b) are mirror-like after mechanical polish.}
  \label{fig4}
\end{figure}

Surface quality is critical for reducing dark current. After dedicated mechanical polishing, the roughness (Ra) of the gun cavity inner surface and cathode plug is lower than 100 nm and 10 nm. The cathode nose and cathode plug of the ZJLAB/SARI VHF gun are mirror-like finish, as depicted in Fig. \ref{fig4}.

The ZJLAB/SARI VHF gun cavity was cleaned in an ultrasonic bath with ultrapure water and rinsed with high pressure water. The gun assembly took place in a Class 100 clean tent, following SHINE particle free assembly procedure. A $\rm CO_{2}$ snow cleaning of the inner surface was carried out. Afterwards, the purified nitrogen gas blowing cleaning was performed.

\section{\label{sec:4}DARK CURRENT MEASUREMENTS}

We built a gun test stand to characterize both high power CW RF and high average current of the ZJLAB/SARI VHF gun,  as shown in Fig. \ref{fig5.png}. The test stand consists of an INFN-type load-lock system, a VHF gun, two solenoids, steer magnets, beam diagnostics, an energy spectrometer, and a beam dump. There are three collimators with hole diameter of 16 mm, 20 mm and 24 mm located at the \#1 profile station, which is situated 1.47 m downstream from the cathode. A Faraday Cup (FC) was installed at the \#3 profile station, located at 2.23 m downstream. 

\begin{figure*}
    \centering
    \includegraphics[width=1\linewidth]{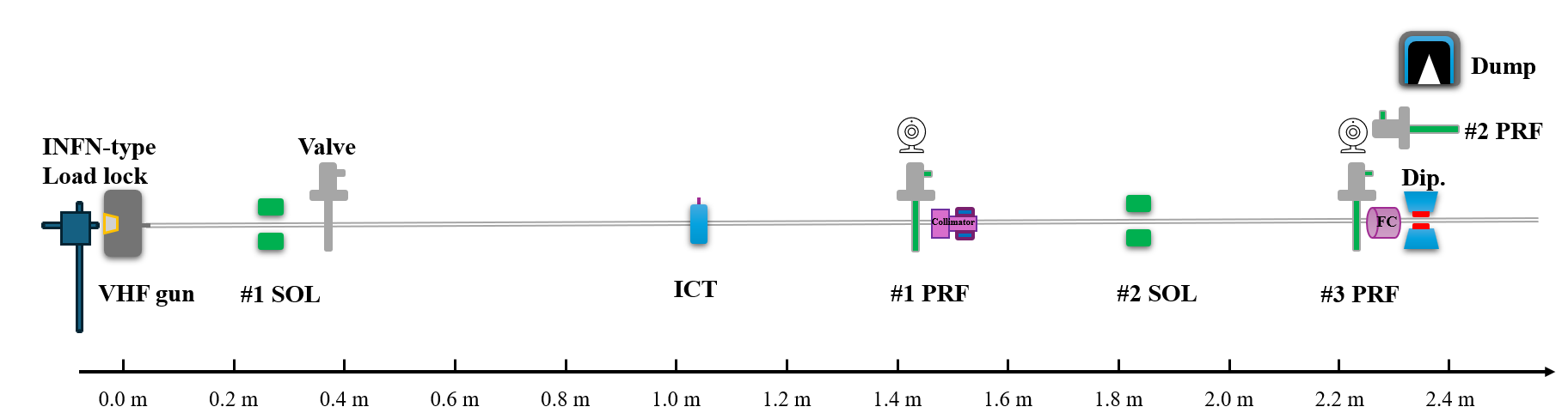}
    \caption{\label{fig5.png}~Layout of the ZJLAB/SARI VHF gun test stand.}
    \label{fig5.png}
\end{figure*}

The dark current is measured using a FC, as illustrated in Fig. \ref{fig5.png}. The typical dark current and RF pulse waveforms using an oscilloscope (Keysight EXR204A) are shown in Fig. \ref{fig6.png}. With an input impedance of 1 M$\Omega$ in the oscilloscope, the voltage of dark current signal is 200 mV, corresponding to the dark current of 200 nA. The noise level of the FC is $\sim$ 1 nA. 

\begin{figure}
    \centering
    \includegraphics[width=8.6cm]{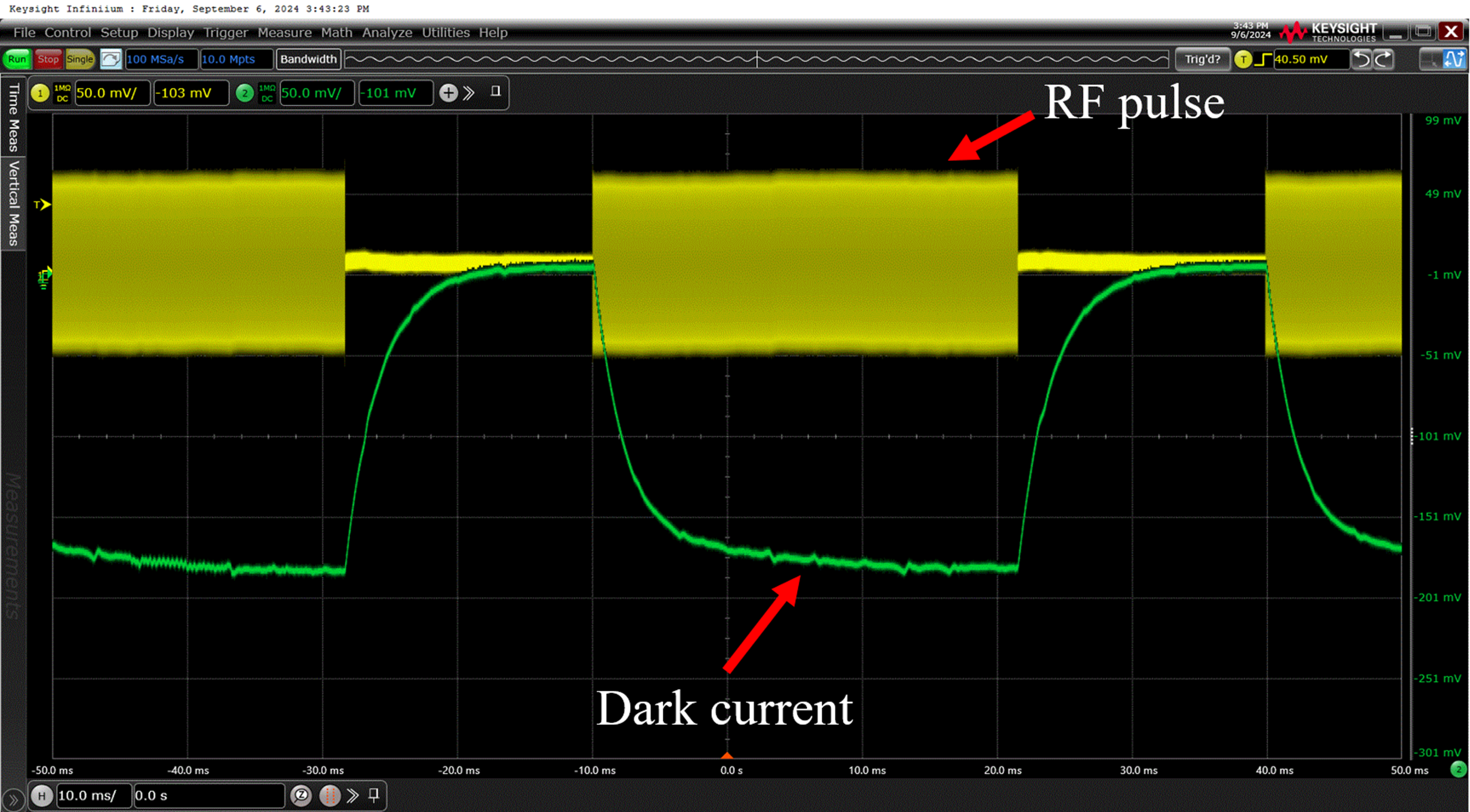}
    \caption{\label{fig6.png}~Typical dark current and RF pulse signals from the FC. The input RF pulse length is 32 ms with a repetition rate of 20 Hz.}
    \label{fig6.png}
\end{figure}

\begin{figure}
    \centering
    \includegraphics[width=6.8cm]{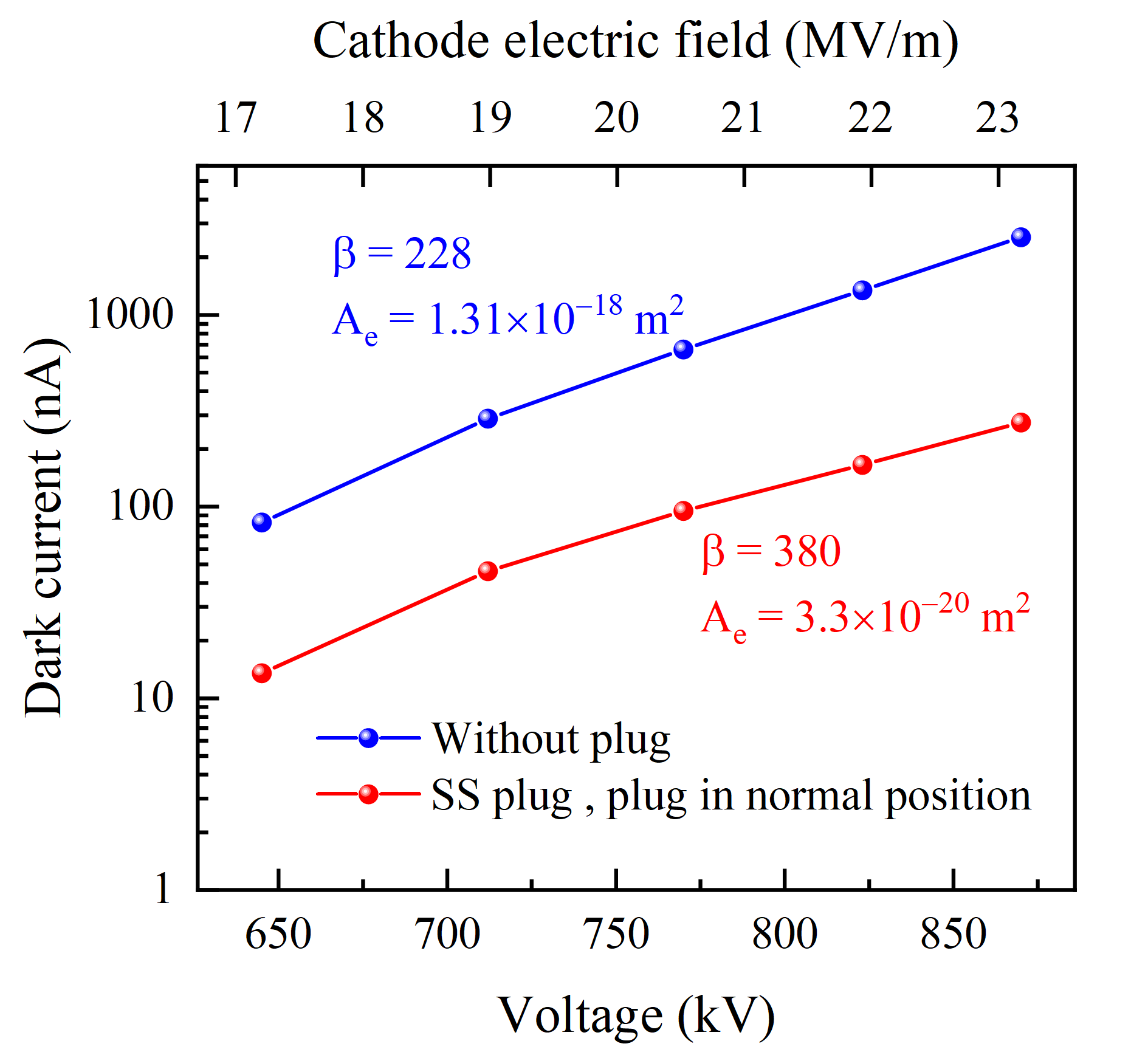}
    \caption{\label{fig7.png}~Dark current as a function of the gun voltage and the cathode electric field. Red dots: SS plug in the normal insertion. Blue dots: plug was pulled out, the dark current was measured at the same RF power with the red dots. The work function of SS is 4.1 eV \cite{2012Experimental}, which is used to fit the parameters $\rm \beta$ and $\rm A_{e}$.}
    \label{fig7.png}
\end{figure}

\begin{figure}
    \centering
    \includegraphics[width=8.4cm]{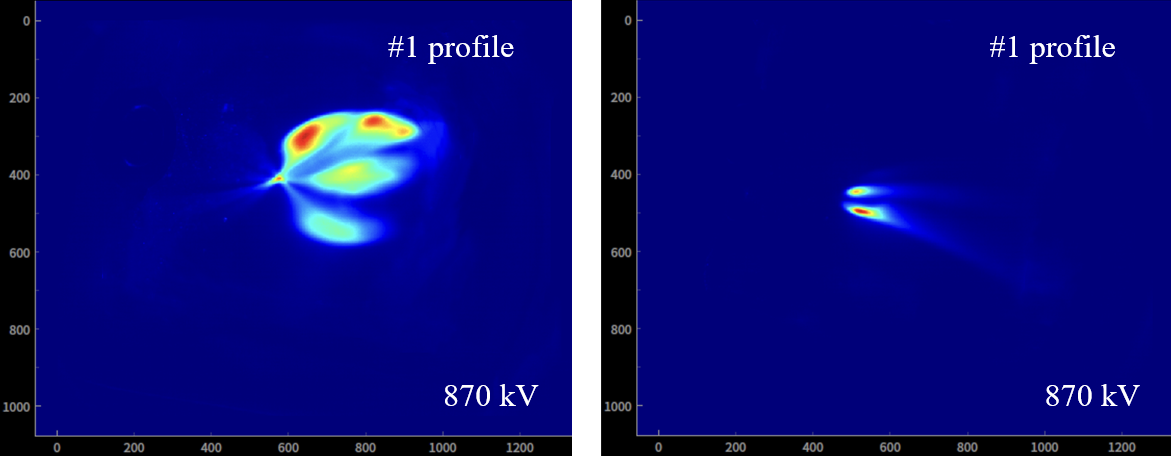}
    \caption{\label{fig8.png}~Dark current imaging at \#1 profile screen with 1\% RF duty factor. Left: plug out, dark current at 870 kV was 2.54 $\mu$A. Right: SS plug inside, dark current at 870 kV was 275 nA. }
    \label{fig8.png}
\end{figure}

Figure \ref{fig7.png} presents the dark current measurements with and without the cathode plug. The cathode plug was made by the SS 316L without $\rm Cs_{2}Te$ material. The measurements were carried out at an RF duty factor of 1\%. The current of the solenoids and steer magnets were adjusted to achieve the maximum dark current value. From the Fowler-Nordheim (F-N) theory \cite{fn1,fn2,fn3}, the average field emission current $I_{F}$ (in A) during one RF period can be described as Eq. \ref{eq1}:

\begin{equation} \begin{aligned} 
I_F &= \frac{5.7 \times10^{-12} \times10^{4.52 \phi^{-0.5}} A_e (\beta E)^{2.5}}{\phi^{1.75}} \\
 & \quad \times \exp\left(-\frac{6.53 \times10^{9} \phi^{1.5}}{\beta E}\right)  
\end{aligned}  \label{eq1}
\end{equation}  

where $\rm \phi$ is the work function of the metal material in eV, $\rm A_{e}$ is the effective emitting area (in $\rm m^2$), $\rm\beta$ is the field enhancement induced by microscopic surface defects such as scratches, protrusions and particulates, $E$ is the surface electric field in V/m. The fitted parameters $\rm\beta$ and $\rm A_{e}$ can be found in Fig. \ref{fig7.png}. With the plug inserted inside, the field enhancement factor $\rm\beta$ is increased but the emission area $\rm A_{e}$ is reduced by two orders of magnitude, resulting in a dark current reduction by $\rm\sim$ 90\%.

The dark current images on the \#1 profile screen, with and without cathode plug, were obtained at the gun voltage of 870 kV, as shown in Fig.  \ref{fig8.png}. After a SS plug without $\rm Cs_{2}Te$ was inserted into the gun, the dark current was reduced from 2.54 $\rm \mu$A to 275 nA. The dark current signals on the profile were also significantly reduced. The momentum of the dark current was similar to that of the photo beams, indicating that the emitters originated from the vicinity of the cathode. Rotating the plug by approximately $180^{\circ}$ resulted in no observable change in the dark current image, supporting the conclusion that the dark current originates from the cathode backplate.

\section{\label{sec:5}TRACKING SIMULATIONS}

\subsection{Tracking simulations with the normal insertion plug}

Based on the experimental results presented in the previous Section \ref{sec:4}, the plug hole region is identified as the primary source of downstream dark current. The RF defocusing field on the plug corner and the focusing field on the plug hole region significantly influence the trajectories of field emitted electrons. We employed CST MWS to calculate the three-dimensional (3D) electromagnetic field in the ZJLAB/SARI gun cavity. The local mesh of the cathode plug vicinity was refined to obtain an accurate electromagnetic field distribution. The 3D fields were imported into ASTRA \cite{ASTRA} to track the field emission trajectories and calculate the transmission ratio (TR) from emission position to the FC. The temporal distribution of the field emission was derived from the F-N formula using the experimental data shown in Fig. \ref{fig7.png}. A Gaussian fit was utilized in ASTRA to define the initial longitudinal distribution of the field emission, as illustrated in Fig. \ref{fig9.png}. In the tracking simulations, the impacts of the space charge force, multipacting inside the gun and geomagnetic field were not considered.

\begin{figure}
    \centering
    \includegraphics[width=6.8cm]{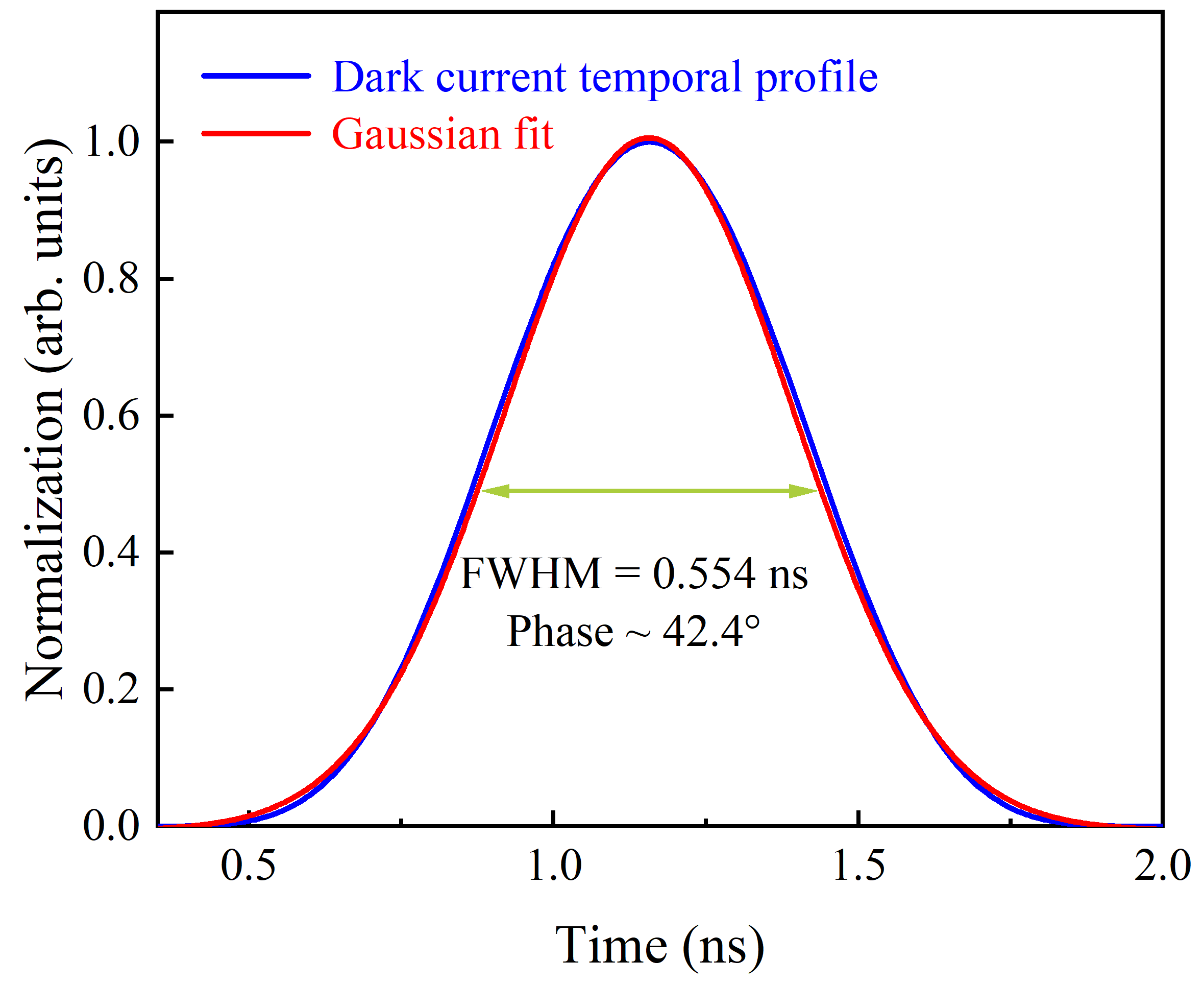}
    \caption{\label{fig9.png}~The temporal distribution of the field emissions in one RF cycle. The FWHM of the Gaussian fit is 0.554 ns which corresponds to $42.4^{\circ}$ phase.}
    \label{fig9.png}
\end{figure}

The dark current TR, shown in Fig. \ref{fig10.png}, was calculated with a \#1 solenoid current of 44 A and \#2 solenoid current of 35 A, corresponding to the maximum dark current captured by the FC at the gun voltage of 750 kV. The emitters from R $\textgreater$ 9 mm region are lost during the transmission. Therefore, surface processing within R $\textless$ 9 mm region are critical. A correlation has been observed between TR and the transverse electric field at the surface. On the plug corner, the surface electric field serves as a defocuing effect resulted in a TR reduction. Conversely, at the hole corner, the electrons experience a focusing effect, which increases TR.

\begin{figure}
    \centering
    \includegraphics[width=8.6cm]{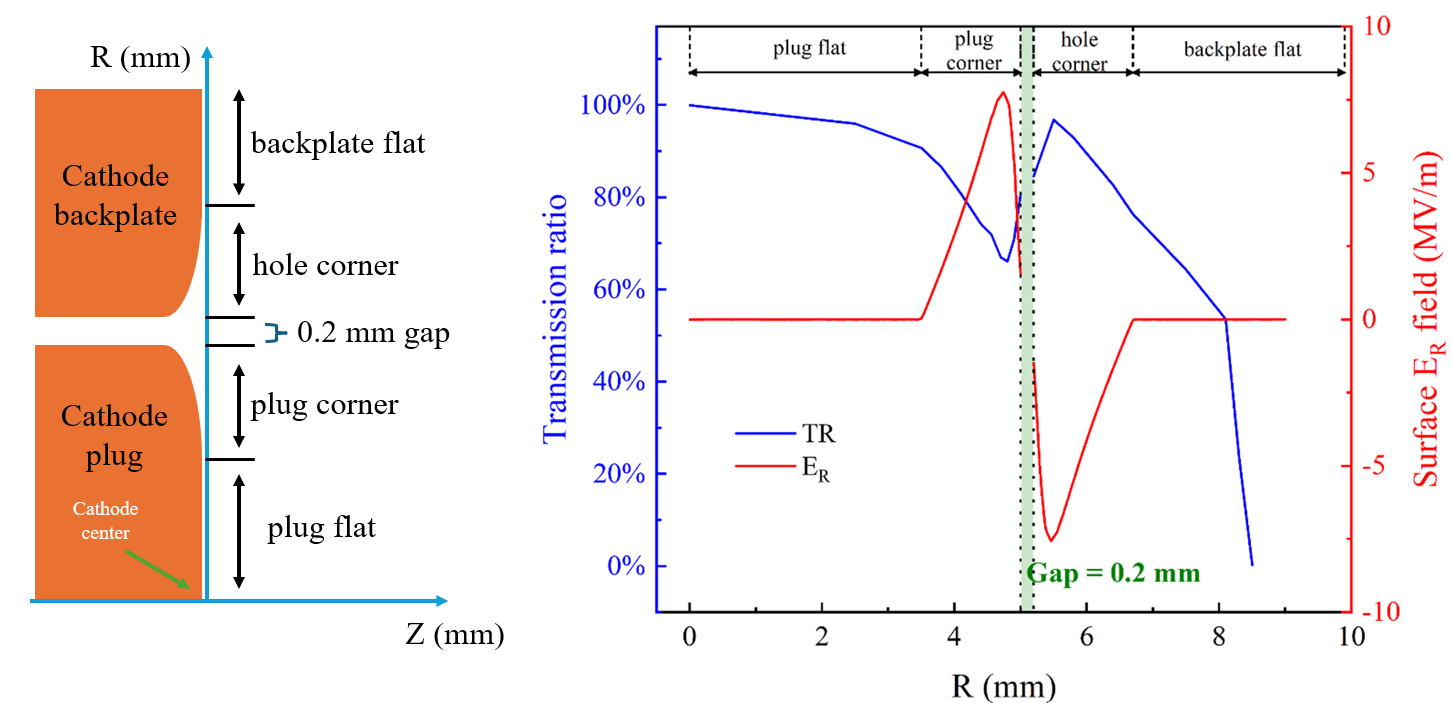}
    \caption{\label{fig10.png}~Left: the cathode vicinity was divided into four regions: plug flat, plug corner, hole corner and backplate flat. Right: Field emission transmission ratio and surface radial electric field as a function of emitted position.}
    \label{fig10.png}
\end{figure}

\subsection{Tracking simulations with various plug insertion depths}	

The concept of an over-inserted plug was proposed by the PITZ group to suppress dark current \cite{39}. An over-inserted plug reduces both the peak electric field and the focusing field at the plug hole. Fig. \ref{fig11} shows the electric fields in the ZJLAB/SARI VHF gun with various insertion depths at a constant RF power of 70 kW. A 1.0 mm over-insertion results in a 17.7\% decrease in the peak electric field on the cathode hole but a 55.4\% increase at the cathode plug corner. In contrast, a retracted plug exhibits an opposite effect on the surface electric field. If the plug is pulled out the cavity, the field on the plug hole corner will be increased by $\sim$ 44.5\% comparing to the plug in the normal insertion.

\begin{figure}
    \centering
    \includegraphics[width=6.8cm]{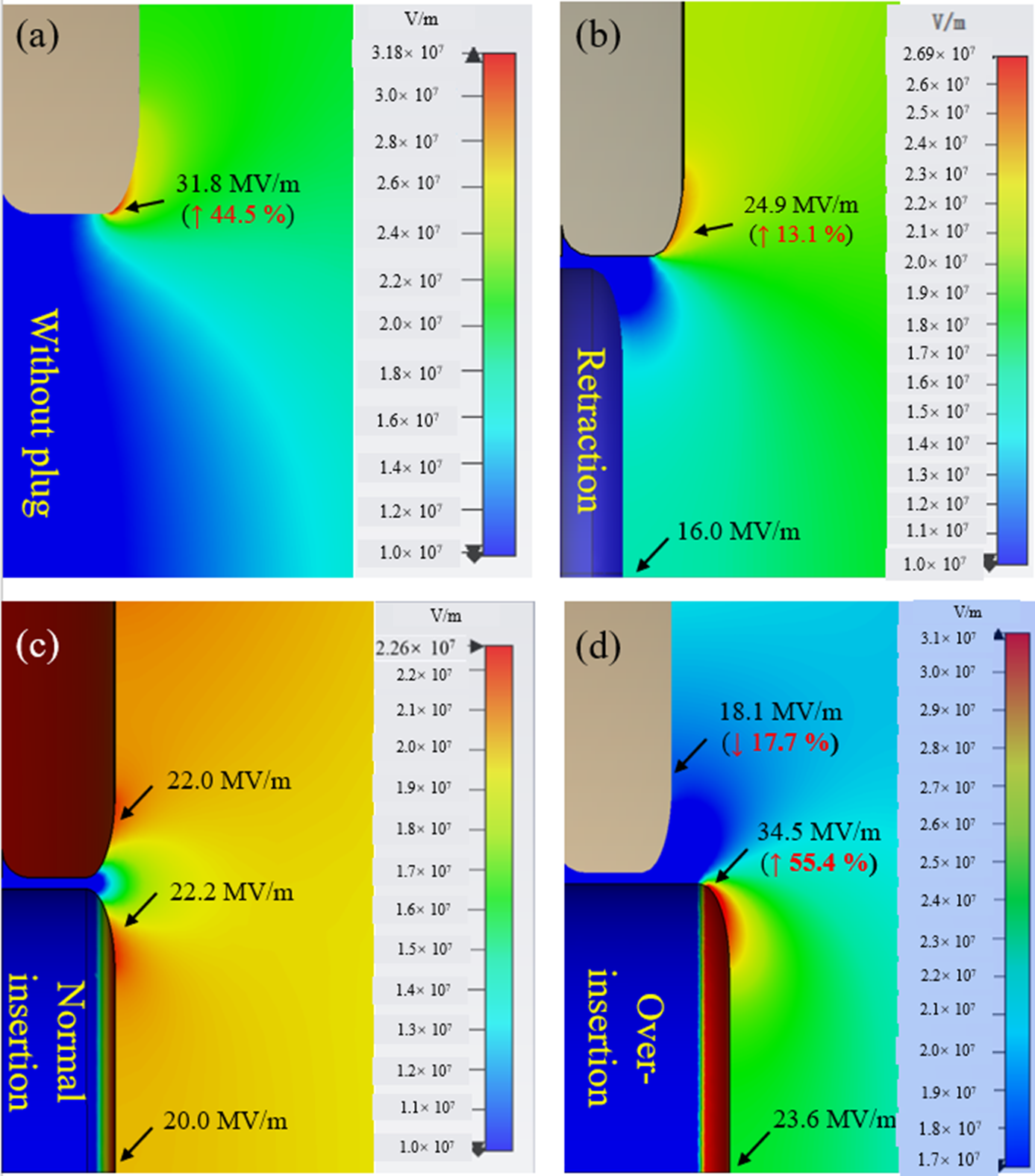}
    \caption{\label{fig11}~Electric fields in the cathode vicinity at 70 kW RF power. (a) plug is out. (b) plug is retracted by 1 mm. (c) plug in normal insertion. (d) plug is over-inserted by 1 mm.}
     \label{fig11}
\end{figure}

\begin{figure}
    \centering
    \includegraphics[width=8.3cm]{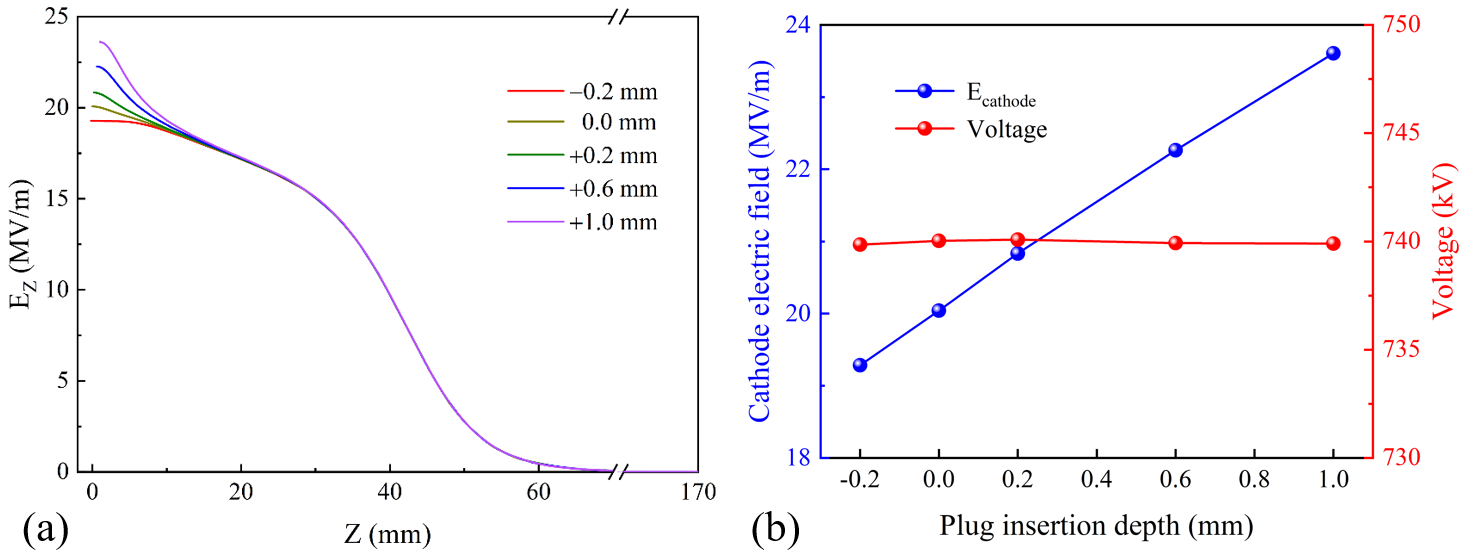}
    \caption{ ~(a) Electric fields along the beam line at 70 kW RF power. (b) Cathode gradient and gun voltage as a function of the plug insertion depth at a constant RF power of 70 kW.}
     \label{fig12}
\end{figure}

The electrical fields along the beam line at various insertion depths are presented in Fig. \ref{fig12}. At 70 kW RF power, the cathode gradient increases from 20 MV/m to 23.6 MV/m with a 1 mm over-insertion, with the voltage change remaining within 0.2 kV.

\begin{figure}
    \centering
    \includegraphics[width=8.3cm]{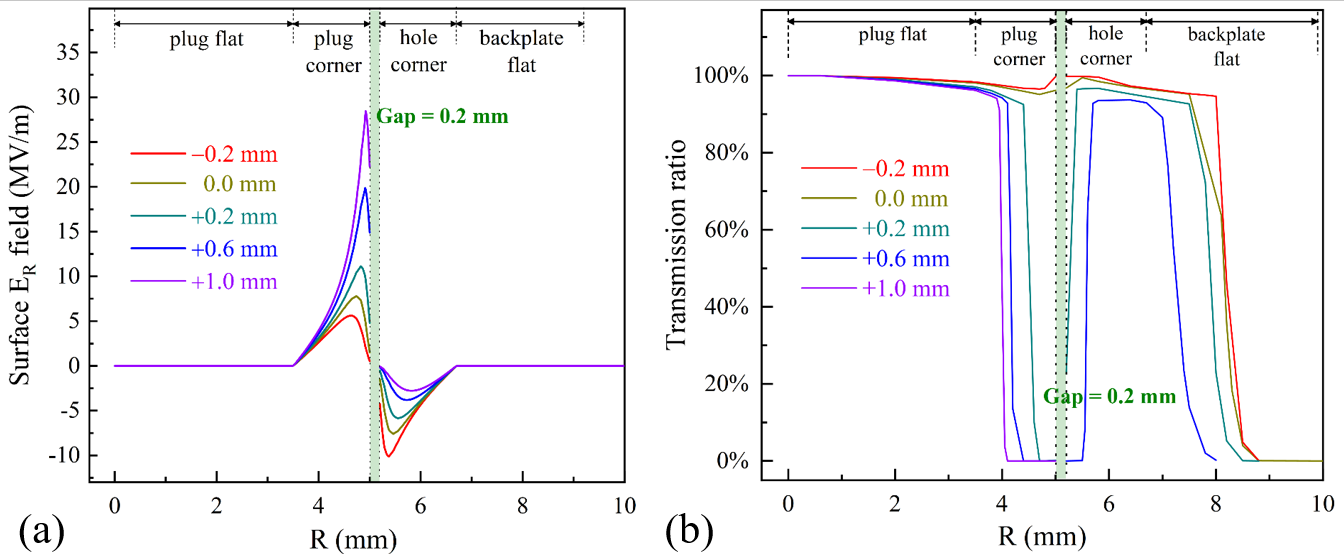}
    \caption{\label{fig13.a.png} \label{fig13.b.png} ~(a) Surface transverse electric field in the cathode vicinity, RF power is 70 kW. (b) Dark current TR in the with five different cathode insertion depths.}
     \label{fig13}
\end{figure}

The field emission tracking simulation results with different insertion depths are presented in Fig. \ref{fig13}. As the cathode insertion depth increases, the high TR region is narrowed for both the plug and the backplate hole. Fig. \ref{fig13.a.png}(a) indicates that an over-inserted plug creates a stronger defocusing field at the plug corner and a weaker focusing field at the hole corner, resulting in a reduction of TR, as shown in Fig. \ref{fig13.b.png}(b). If the plug is over-inserted by more than 1 mm, electrons from the backplate cannot escape from the gun, making the plug the sole source of dark current.

With an over-inserted plug, the electric field on the cathode backplate is reduced and the high TR region is narrowed. If the dark current is dominated by the cathode backplate, and the plug is well polished, employing an over-inserted plug might be effective to reduce dark current.

\begin{figure}[htbp]
	\centering
	\subfigure[~]{\label{fig14a.png}\includegraphics[width=8.6cm]{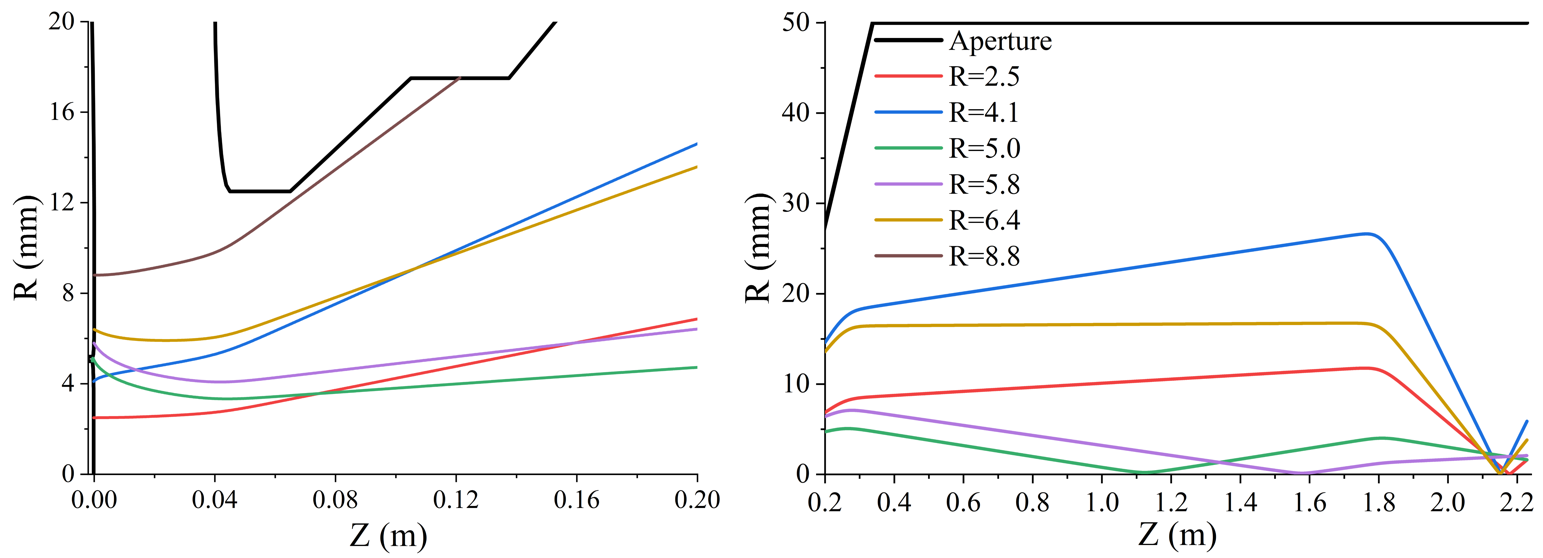}}
	\subfigure[~]{\label{fig14b.png}\includegraphics[width=8.6cm]{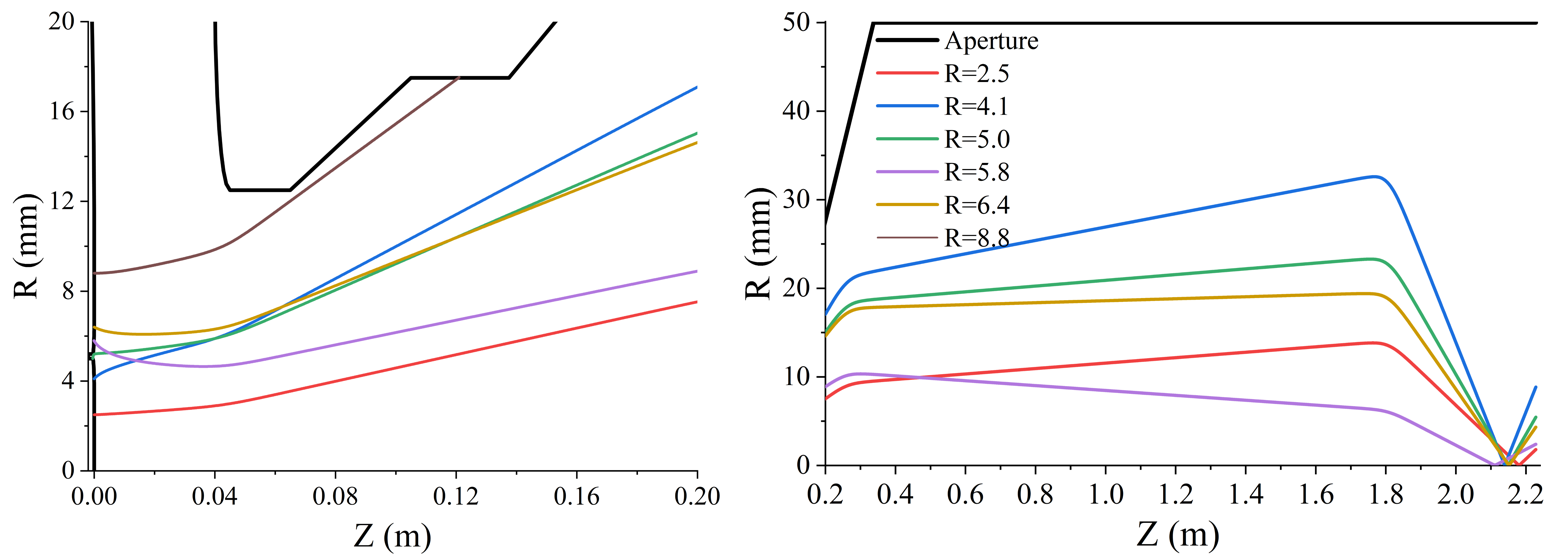}}
 \subfigure[~]{\label{fig14d.png}\includegraphics[width=8.6cm]{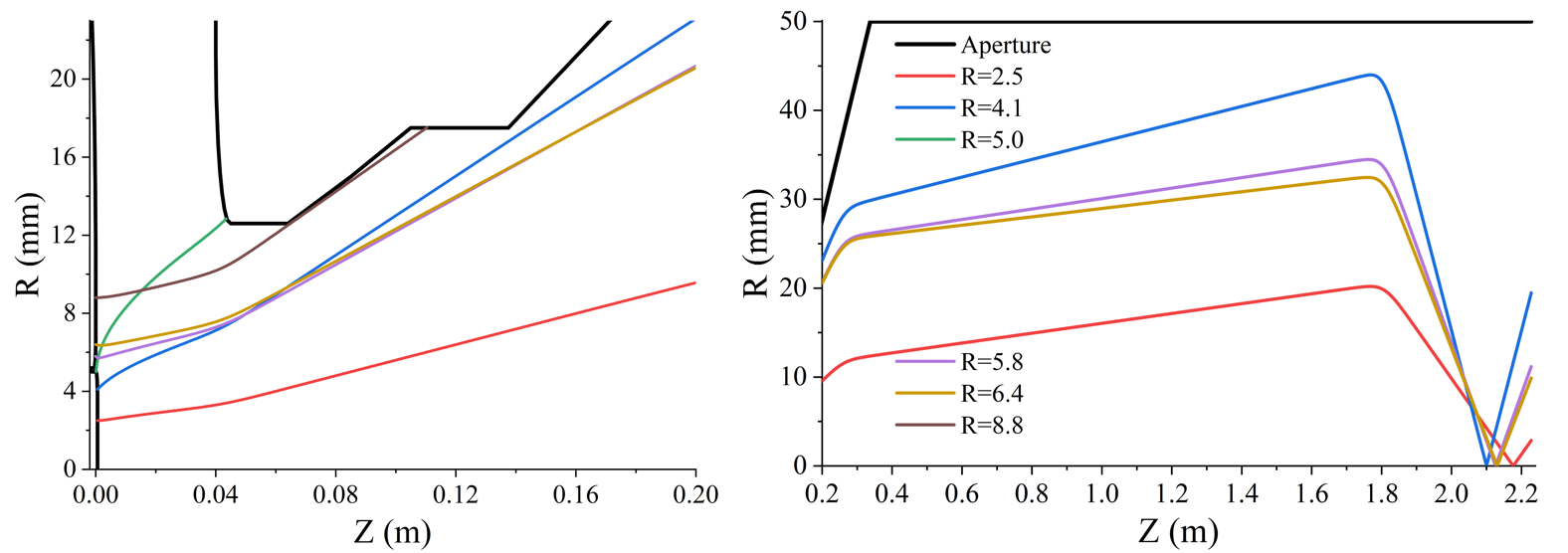}}
 \subfigure[~]{\label{fig14e.png}\includegraphics[width=8.6cm]{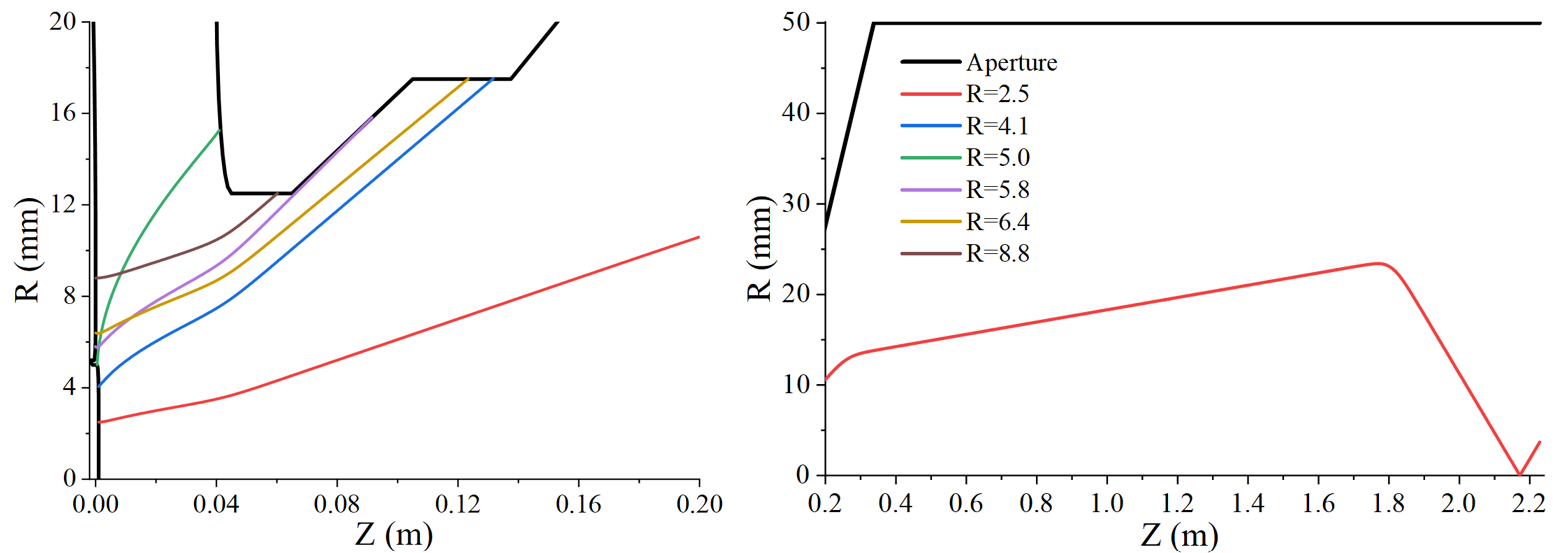}}
	\caption{~The emission trajectories of some typical points. The left plots display the electron trajectories in VHF gun. The right plots show the electron trajectories in the downstream beamline. (a) 0.2 mm retraction, (b) normal insertion, (c) 0.6 mm over-insertion, (d) 1.0 mm over-insertion.}
 \label{fig:14}
\end{figure}

Figure. \ref{fig:14} presents the particle trajectories in the ZJLAB/SARI VHF gun test stand, tracking several typical emission positions in the cathode vicinity with the plug at various insertion depths. The electrons emitted in the gun phase of $90^{\circ}$ (maximum surface electric field) were tracked. The solenoids were set at the similar currents as depicted in Fig. \ref{fig10.png}. As the insertion depth increases, a broader range of field emitted electrons are trapped within the gun body or lost along the beam line pipe.

\section{\label{sec:6}~DARK CURRENT MEASUREMENTS WITH OVER-INSERTED PLUGS}	

We fabricated five SS cathode plugs (no $\rm Cs_{2}Te$) with different over-insertion depths ranging from 0.3 mm to 1.5 mm. The plugs have a surface roughness of less than 10 nm. The gun frequency before and after the plug insertion was measured using the low level RF system (LLRF). The frequency change was compared to simulation results to verify the plug insertion depth accurately.

\begin{figure}
    \centering
    \includegraphics[width=6.8cm]{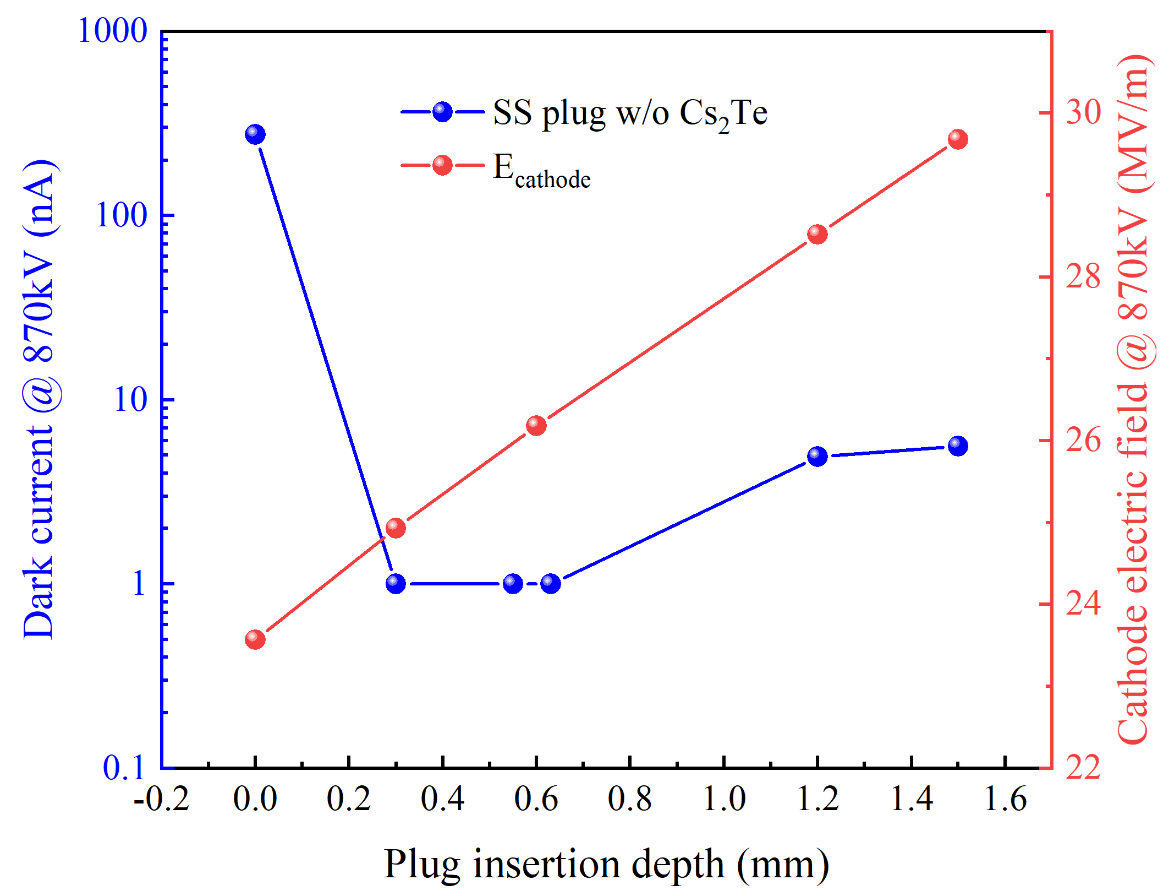}
    \caption{\label{fig15.png}~Dark current measurements (blue dots) and cathode electric field values (red dots) as a function of plug insertion depth at a gun voltage of 870 kV.}
    \label{fig:enter-label}
\end{figure}

\begin{figure}
    \centering
    \includegraphics[width=8cm]{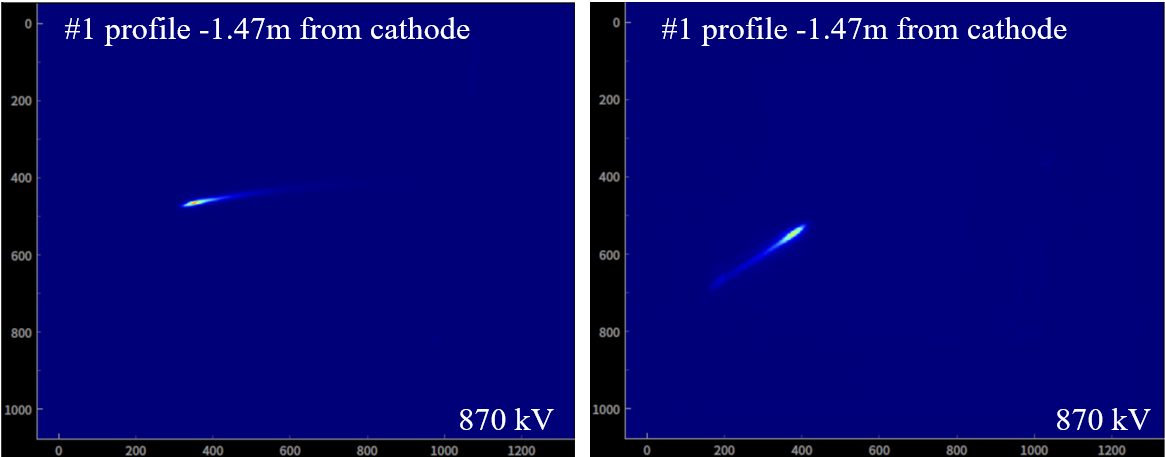}
    \caption{\label{fig17.png}~Dark current images at the gun voltage of 870 kV with a 1.2 mm over-inserted plug. Left: before plug rotation, Right: after plug rotation.}
    \label{fig17}
\end{figure}

Figure. \ref{fig15.png} summarizes the dark current measurements at the ZJLAB/SARI VHF gun test stand, with a constant RF power of 100 kW and a gun voltage of $\rm\sim$ 870 kV. When the cathode plug was over-inserted by 0.3 to 0.6 mm, the dark current fell within the noise level of the FC electronics ($\rm\sim$ 1 nA). The dark current was not detected at the \#1 profile screen with a sufficiently long camera exposure time and an adequately high camera gain. However, when the insertion depth exceeded 1.2 mm, dark current was observed on the screen. The dark current image rotated with the rotation of the cathode plug, as shown in Fig. \ref{fig17}, indicating that the primary source of dark current shifted from the previous backplate hole to the plug itself.

The dark current image with a normal inserted plug in Section \ref{sec:4} indicates that several emitters from the cathode backplate are the dark current source. The precise positions of the emitters are unknown. With a 0.6 mm over-inserted plug, the electric field on the plug backplate was reduced by 10\%. However, the dark current was dramatically reduced from 275 nA to FC noise level ($\rm\sim$ 1 nA) and disappeared on the screen, which indicated the trajectory change is the major reason for the dark current reduction.

\begin{figure}
	\centering
	\subfigure[~]{\label{fig18a.png}\includegraphics[width=6.8cm]{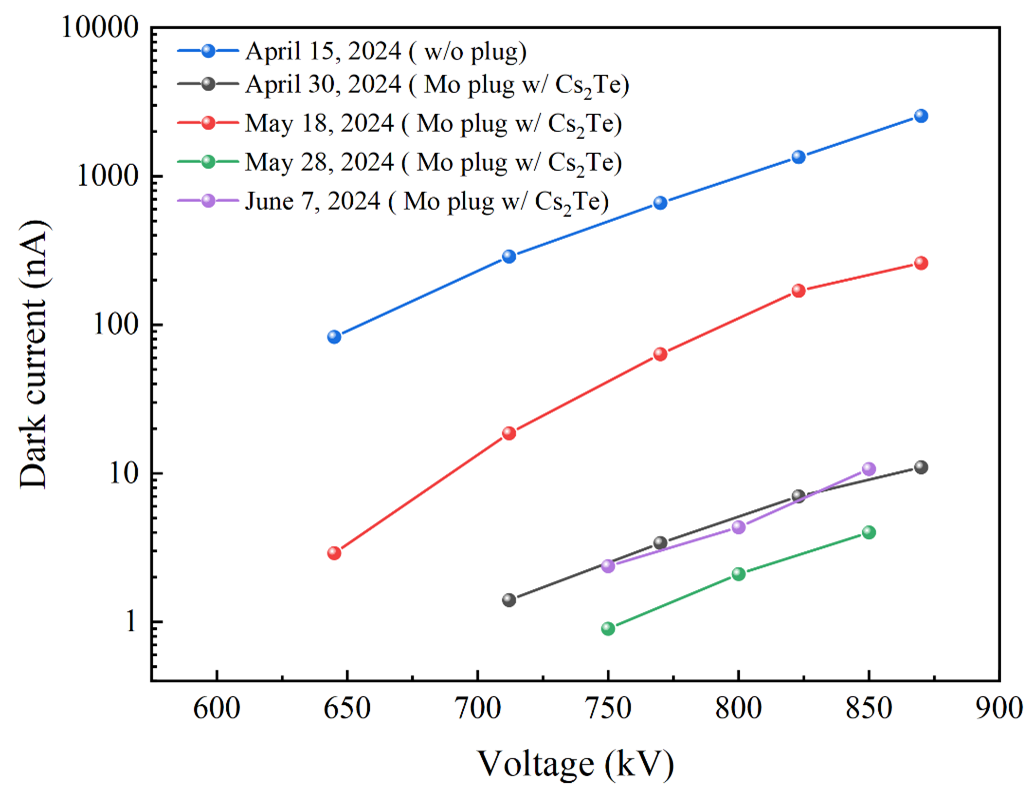}}
	\subfigure[~]{\label{fig18b.png}\includegraphics[width=4.6cm]{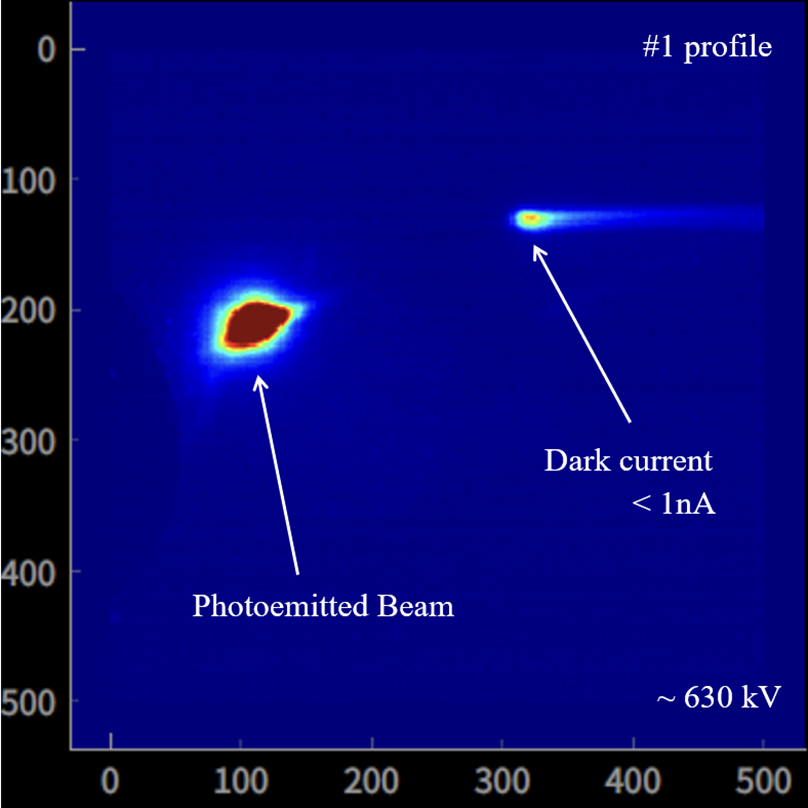}}
	\caption{~ (a) Dark current evolution along the running time. (b) Image of the photon beam and dark current on the \#1 profile.}
\end{figure}

A 0.6 mm over-inserted molybdenum (Mo) plug with $\rm Cs_{2}Te$ layer was measured in the ZJLAB/SARI VHF gun. Fig. \ref{fig18a.png} shows the dark current evolution. Initially, the dark current was 2.54 $\rm \mu$A without plug at the gun voltage of 870 kV. Then it was reduced to 11.0 nA after the plug insertion on April $\rm 30^{th}$, 2024. Throughout the laser alignment process during the $\rm 2^{nd}$ week of May, the plug was rotated several times, and the dark current at 870 kV increased to 260 nA on May $\rm 18^{th}$. Two potential reasons could explain this increase: first, the over-insertion depth of the plug may have changed during laser alignment; second, the friction during the rotation of the cathode plug may have led to the generation of particulates, introducing additional emitters. On May $\rm 28^{th}$, the plug was extracted and re-inserted into the gun again, with the insertion depth confirmed via frequency measurement. The dark current was 4.0 nA at 850 kV. The dark current on June $\rm 7^{th}$ was 10.7 nA after $\rm\sim$ 1 week laser illumination which generated $\rm\sim$ 10 C total charge. Fig. \ref{fig18b.png} presents the image of dark current and photon beam at the \#1 profile taken on May $\rm 30^{th}$. Notably, a significant increase in dark current was not observed during the beam commissioning phase. Moreover, we monitored the dark currents during 72 hours of uninterrupted high-power CW operation (50 kW $\rm\&$ 630 kV) without laser illumination, and the dark current remained unchanged throughout this period.

\begin{figure}[htpb!]
    \centering
    \includegraphics[width=8cm]{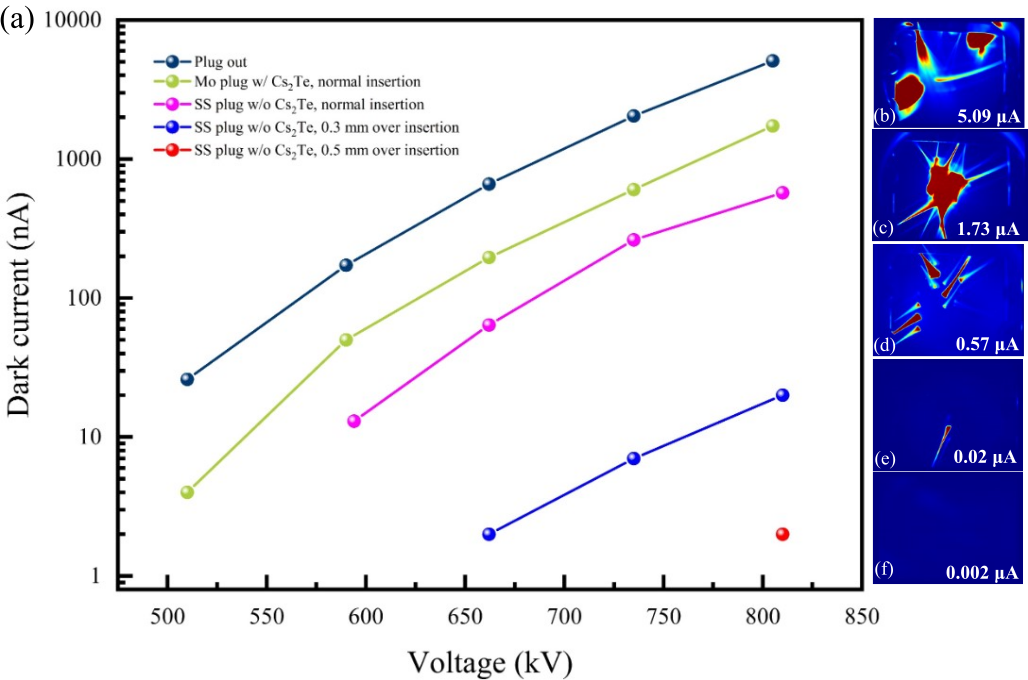}
    \caption{\label{fig19.png}~(a) : SHINE gun dark current vs. voltage. (b)$\sim$(f) : dark curren images at 810 kV voltage. (b) plug out; (c) Mo plug with $\rm Cs_{2}Te$ in normal insertion; (d) SS plug without $\rm Cs_{2}Te$ in normal insertion; (e) SS plug without $\rm Cs_{2}Te$, 0.3 mm over-insertion; (f) SS plug without $\rm Cs_{2}Te$, 0.5 mm over-insertion.}
    \label{fig19}
\end{figure}

The VHF gun installed in SHINE injector experienced an increase in dark current during operation. The dark current measurements are shown in Fig. \ref{fig19}.At the gun voltage of 810 kV, the dark current without plug was 5.09 $\rm \mu$A. The dark current of a normal inserted Mo plug with $\rm Cs_{2}Te$ was 1.73 $\rm \mu$A. Three SS plug with different insertion depths were tested in the SHINE VHF gun. The dark current was 570 nA with a normal inserted SS plug. This value decreased to 28 nA with a 0.3 mm over-inserted SS plug and further reduced to 2 nA with a 0.5 mm over-inserted SS plug. The dark current was reduced by more than two orders of magnitude with the over-inserted plugs. As shown in Fig. \ref{fig19}(e), emitters disappeared with the 0.5 mm over-inserted plug. The dark current from cathode backplate was completely blocked. 

The ASTRA simulations were performed to compare the beam properties of normal plug insertion and 0.5 mm over-insertion. For the normal inserted plug, beam dynamics optimization of the SHINE injector was carried out based on a cathode gradient of 25 MV/m and a gun voltage of 730 kV \cite{33}. The optimized parameters were then applied to estimate the beam characteristics emitted from the 0.5 mm over-inserted plug. The cathode gradient was increased to 27.6 MV/m, while the gun voltage remained constant with a fixed input RF power. The beam performance was comparable to that of the normal plug insertion.

 Additionally, we employed the Multi-Objective Genetic Algorithm (MOGA) to further optimize the 0.5 mm over-insertion configuration, utilizing the current beamline layout of the SHINE injector.Nine parameters were tuned to minimize both the emittance and bunch length, including laser transverse and longitudinal distributions, gun phase, buncher voltage and phase, two solenoid currents, the amplitude of the single 9-cell SRF cryomodule, and the amplitude of the first SRF cavity in the eight 9-cell SRF cryomodule. The optimized parameters obtained were similar to those in the normal plug insertion situation. The beam performance for the 0.5 mm over-insertion was also similar to that of the normal plug insertion.

\section{~CONCLUSIONS AND DISCUSSIONS}		

A normal conducting VHF band gun was successfully developed and commissioned with the collaboration between ZJLAB and SARI from 2022 to 2024. The systematic study of the dark current is presented in this paper. In order to mitigate the dark current of the ZJLAB/SARI VHF gun, the elliptical profile was adopted in the cathode vicinity to reduce the surface electric field. The criteria for cavity surface polishing, cleaning, and gun assembly were established. The SS was chosen as the material for the backplate hole. 

The dark current emission of the VHF gun is dominated by the cathode hole corner. By using a properly over-inserted plug, both the electric field intensity and dark current transmission ratio are reduced. The experiments of the ZJLAB/SARI VHF gun demonstrated that the dark current was reduced from 275 nA (with a normal inserted plug) to FC noise level ($\rm\sim$ 1 nA) with a 0.3-0.6 mm over-inserted plug at the gun voltage of 870 kV. The dark current image revealed that the emitters from the cathode backplate were fully blocked, and the emitters from the plug itself were the only sources of dark current. No significant increase in dark current was observed after approximately one month high RF power operation and beam commissioning. 

Additionally, the application of a 0.5 mm over-inserted plug in the SHINE VHF gun resulted in a dark current reduction from 570 nA to 2 nA at 810 kV voltage. The beam performance remained similar according to the ASTRA optimizations.

For the VHF guns, in which dark current is dominated by the cathode backplate, an over-inserted plug is effective in reducing dark current. A retracted plug, which is widely used in the SRF guns for additional RF focusing might lead to higher dark current emission in VHF guns. The optimal insertion depth should be determined through experiments. Based on our experiences with various SS plug insertions, the dark current decreases initially and then increases as the insertion depth increases, indicating that the dominance of dark current changes from the backplate to the plug. It has been demonstrated that a 0.3-0.6 mm over-inserted plug can reduce the dark current by more than two orders of magnitude. Surface polishing of the plug is crucial for controlling the dark current increase when the surface electric field is enhanced. The dark current is highly sensitive to the plug position in the gun. It was observed that the position of the cathode plug is not always the same after repeated insertions. Frequency measurements before and after plug insertion are essential for ensuring the insertion depth.

\section{~acknowledgments}

The authors would like to acknowledge Y. Du, L. Zheng from Tsinghua University and J. Shao from Institute of Advanced Science Facilities, Shenzhen (IASF) for the fruitful discussions. This work was sponsored by Shanghai Rising-Star Program (22QA1411900), the CAS Project for Young Scientists in Basic Research (YSBR-042), the National Natural Science Foundation of China (12125508, 11935020), and Shanghai Pilot Program for Basic Research-Chinese Academy of Sciences, Shanghai Branch (JCYJ-SHFY-2021-010).

\nocite{*}

\bibliography{reference}

\end{document}